\documentclass[conference]{IEEEtran}
\IEEEoverridecommandlockouts

\usepackage{cite}
\usepackage{amsmath,amssymb,amsfonts}
\usepackage{algorithmic}
\usepackage{graphicx}
\usepackage{textcomp}
\usepackage{xcolor}
\def\BibTeX{{\rm B\kern-.05em{\sc i\kern-.025em b}\kern-.08em
    T\kern-.1667em\lower.7ex\hbox{E}\kern-.125emX}}
\begin{document}

\title{DCCRGAN: Deep Complex Convolution Recurrent Generator Adversarial Network for Speech Enhancement}

\author{\IEEEauthorblockN{1\textsuperscript{st} Huixiang Huang}
\IEEEauthorblockA{\textit{ZheJiang Dahua Technology CO.,LTD } \\
Hangzhou, China \\
huixianghuang@yeah.net}
\and
\IEEEauthorblockN{2\textsuperscript{nd} Renjie Wu}
\IEEEauthorblockA{\textit{ZheJiang Dahua Technology CO.,LTD } \\
Hangzhou, China \\
garrywrj@gmail.com}
\and
\IEEEauthorblockN{3\textsuperscript{rd} Jingbiao Huang}
\IEEEauthorblockA{\textit{ZheJiang Dahua Technology CO.,LTD } \\
Hangzhou, China \\
huang\_jingbiao@dahuatech.com}
\and
\IEEEauthorblockN{4\textsuperscript{th} Jucai Lin}
\IEEEauthorblockA{\textit{ZheJiang Dahua Technology CO.,LTD } \\
Hangzhou, China \\
lin\_jucai@dahuatech.com}
\and
\IEEEauthorblockN{5\textsuperscript{th} Jun Yin}
\IEEEauthorblockA{\textit{ZheJiang Dahua Technology CO.,LTD } \\
Hangzhou, China \\
yin\_jun@dahuatech.com}
}

\maketitle

\begin{abstract}
Generative adversarial network (GAN) still exists some problems in dealing with speech enhancement (SE) task. Some GAN-based systems adopt the same structure from Pixel-to-Pixel directly without special optimization. The importance of the generator network has not been fully explored. Other related researches change the generator network but operate in the time-frequency domain, which ignores the phase mismatch problem. In order to solve these problems, a deep complex convolution recurrent GAN (DCCRGAN) structure is proposed in this paper. The complex module builds the correlation between magnitude and phase of the waveform and has been proved to be effective. The proposed structure is trained in an end-to-end way. Different LSTM layers are used in the generator network to sufficiently explore the speech enhancement performance of DCCRGAN. The experimental results confirm that the proposed DCCRGAN outperforms the state-of-the-art GAN-based SE systems.
\end{abstract}


\section{Introduction}

Speech enhancement (SE) aims to improve the quality and intelligibility of noisy speech via removing noise \cite{book}, which can be used as a front-end module of automatic speech recognition (ASR) \cite{80901} or hearing aids \cite{2005Spectral}. Traditional speech enhancement methods include spectral subtraction \cite{1163209}, Wiener filtering \cite{1163086}, statistical model \cite{168664} and subspace method \cite{397090}.

Recently, speech enhancement based on deep neural network (DNN) has significantly advanced the state-of-the-art (SOTA) performance and achieves promising result in speech separation \cite{6853860,7194774,6887314} and speech dereverberation \cite{6854479}. There are two mainly training targets on DNN-based SE system, i.e., masking-based and mapping-based \cite{8369155}. Both training targets modify the noisy speech on the time-frequency (T-F) domain and resynthesize the estimated waveform by simply applying the original phase of the noisy speech. This imposes an upper bound on the performance of speech enhancement because the speech quality can be significantly improved when clean phase spectrum is known according to the related work \cite{PALIWAL2011465}.

Generative adversarial network (GAN) \cite{NIPS2014_5ca3e9b1} has achieved a remarkable breakthrough in the field of image generation, and has been explored for speech enhancement \cite{DBLP:conf/interspeech/PascualBS17,8462068,8462581}. SEGAN \cite{DBLP:conf/interspeech/PascualBS17} is the first approach to apply GAN to SE task, which models a mapping between clean waveform and noisy waveform in an end-to-end way. Because of the unstable training process, other GAN-based systems utilize the other objective function to stabilize the training process, such as WGAN \cite{8706647}, SERGAN \cite{8683799}. All of these GAN-based SE systems apply the U-Net architecture in the generator network from image-to-image translation \cite{DBLP:conf/cvpr/IsolaZZE17} directly. The importance of the generator network on SE task has not been fully explored. Another mainstream for modifying the GAN-based systems is changing the generator network, such as MMSE-GAN \cite{8462068}, CRGAN \cite{DBLP:conf/interspeech/ZhangDSWSZSL20}. These algorithms operate in the T-F domain rather than the time-domain, which ignores the phase mismatch between the clean waveform and noisy waveform.

In this paper, we integrate the deep complex convolution recurrent network (DCCRN) \cite{DBLP:conf/interspeech/HuLLXZFWZX20} into GAN-based system and propose the DCCRGAN to solve the SE task in an end-to-end way. DCCRN combines the advantages of DCUNET \cite{DBLP:conf/iclr/ChoiKHKHL19} and CRN \cite{DBLP:conf/interspeech/TanW18}, and has been proved to be effective in the deep noise suppression challenge. DCCRGAN consists of an encoder-decoder structure and multiple LSTM layers between encoder and decoder. The complex-value operation in the DCCRGAN simulates the correlation between magnitude and phase of speech, which can train the complex target effectively. Different LSTM layers are trained to fully evaluate the impact of generator network on speech enhancement, in addition to determining a better performance for the DCCRGAN structure. We compare the proposed model with several GAN-based systems. The results show that the DCCRGAN outperforms previous models under the same objective function. 

The paper is organized as follows. Section 2 introduces the GAN-based speech enhancement. Next, we describe the proposed model architecture in section 3. The experiment setup and the result is reported in section 4, and section 5 ends up the paper with some conclusions.

\section{GAN-BASED SPEECH ENHANCEMENT}

GAN consists of a generator (G) network and a discriminator (D) network. G learns a map between samples \textbf{x} from some prior distribution $\mathcal X$ and samples \textbf{y} from another distribution $\mathcal Y$. D is a binary classifier that is trained to classify the samples coming from $\mathcal Y$ as real, meanwhile the generated samples coming from G as fake. This is a minimax game between G and D with the objective function:
\begin{equation}
\begin{split}
\underset{\text{G}}{\min}\underset{\text{D}}{\max}\text{V}\left(\text{D},\text{G}\right)=\mathbb{E}_{\mathbf{y}\sim p_{\mathbf{y}}\left(\mathbf{y}\right)}\left[\log \left(\text{D}\left(\mathbf{y}\right)\right)\right]
\\
+\mathbb{E}_{\mathbf{x}\sim p_{\mathbf{x}}\left(\mathbf{x}\right)}\left[\log \left(1-\text{D}\left(\text{G}\left(\mathbf{x}\right)\right)\right)\right].
\end{split}
\end{equation}

\begin{figure}[t]
\centerline{\includegraphics[width=8cm]{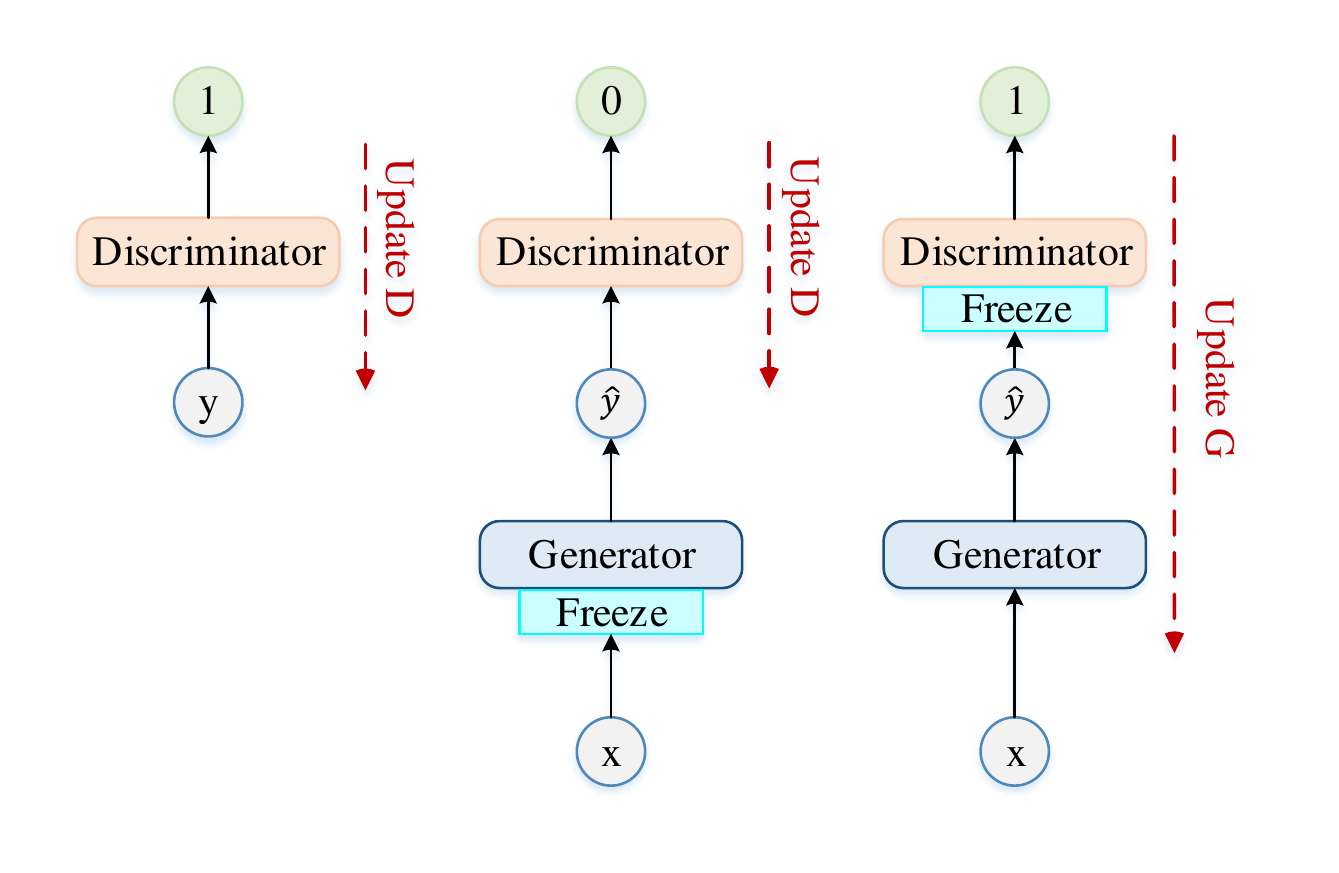}}
\caption{GAN training process.}
\label{fig1}
\end{figure}

As we can see from Fig. \ref{fig1}, GAN’s training process is the repetition of the following three steps:\\
1. Fix the parameters of G, update D such that samples \textbf{y}  are classified as real.\\
2. Fix the parameters of G, update D such that generated samples from G are classified as fake.\\
3. Fix the parameters of D, update G such that generated samples from G are classified as real.

A typical GAN-based end-to-end SE system generally works on conditional GAN (cGAN) \cite{2014Conditional}. G receives the noisy waveform \textbf{x}  such that the generated waveform is closer to the clean waveform  \textbf{y}. D is conditioned using the noisy waveform, so that the input is the clean waveform or the generated waveform concatenating with the noisy waveform, respectively. The relativistic GAN \cite{DBLP:conf/iclr/Jolicoeur-Martineau19} argues that the probability of real samples being real should be decrease as the probability of fake samples being real increase. The relativistic objective function is adopted in SERGAN \cite{8683799} and is proved to be more stable. The relativistic GAN loss:
\begin{equation}
\begin{aligned}
\mathcal{L}_G&=-\mathbb{E}_{\left(\mathbf{x},\mathbf{y}\right)\sim \left(\mathcal{X},\mathcal{Y}\right)}\left[\log\left(\sigma\left(\text{D}\left(\text{G}\left(\mathbf{x}\right),\mathbf{x}\right)-\text{D}\left(\mathbf{y},\mathbf{x}\right)\right)\right)\right]
\\
\mathcal{L}_D&=-\mathbb{E}_{\left(\mathbf{x},\mathbf{y}\right)\sim \left(\mathcal{X},\mathcal{Y}\right)}\left[\log\left(\sigma\left(\text{D}\left(\mathbf{y},\mathbf{x}\right)-\text{D}\left(\text{G}\left(\mathbf{x}\right),\mathbf{x}\right)\right)\right)\right],
\end{aligned}
\end{equation}
where $\sigma$ is sigmoid activation funciton. However, it has a high variance that G influences D as mention in \cite{8683799}, a relativistic average loss is proposed:
\begin{equation}
\begin{aligned}
\mathcal{L}_G&=-\mathbb{E}_{\mathbf{x}\sim \mathcal{X}}\left[\log\left(\overline{\text{D}}_\mathbf{x}\right)\right]-\mathbb{E}_{\mathbf{y}\sim \mathcal{Y}}\left[\log\left(1-\overline{\text{D}}_\mathbf{y}\right)\right]
\\
\mathcal{L}_D&=-\mathbb{E}_{\mathbf{y}\sim \mathcal{Y}}\left[\log\left(\overline{\text{D}}_\mathbf{y}\right)\right]-\mathbb{E}_{\mathbf{x}\sim \mathcal{X}}\left[\log\left(1-\overline{\text{D}}_\mathbf{x}\right)\right],
\end{aligned}
\end{equation}
where $\overline{\text{D}}_\mathbf{y}=\sigma\left(\text{D}\left(\mathbf{y},\mathbf{x}\right)-\mathbb{E}_{\mathbf{x}\sim \mathcal{X}}\left[\text{D}\left(\text{G}\left(\mathbf{x}\right),\mathbf{x}\right)\right]\right)$ and\\
$\overline{\text{D}}_\mathbf{x}=\sigma\left(\text{D}\left(\text{G}\left(\mathbf{x}\right),\mathbf{x}\right)-\mathbb{E}_{\mathbf{y}\sim \mathcal{Y}}\left[\text{D}\left(\mathbf{y},\mathbf{x}\right)\right]\right)$.

\section{PROPOSED DCCRGAN}

\begin{figure}[t]
\centerline{\includegraphics[width=10cm]{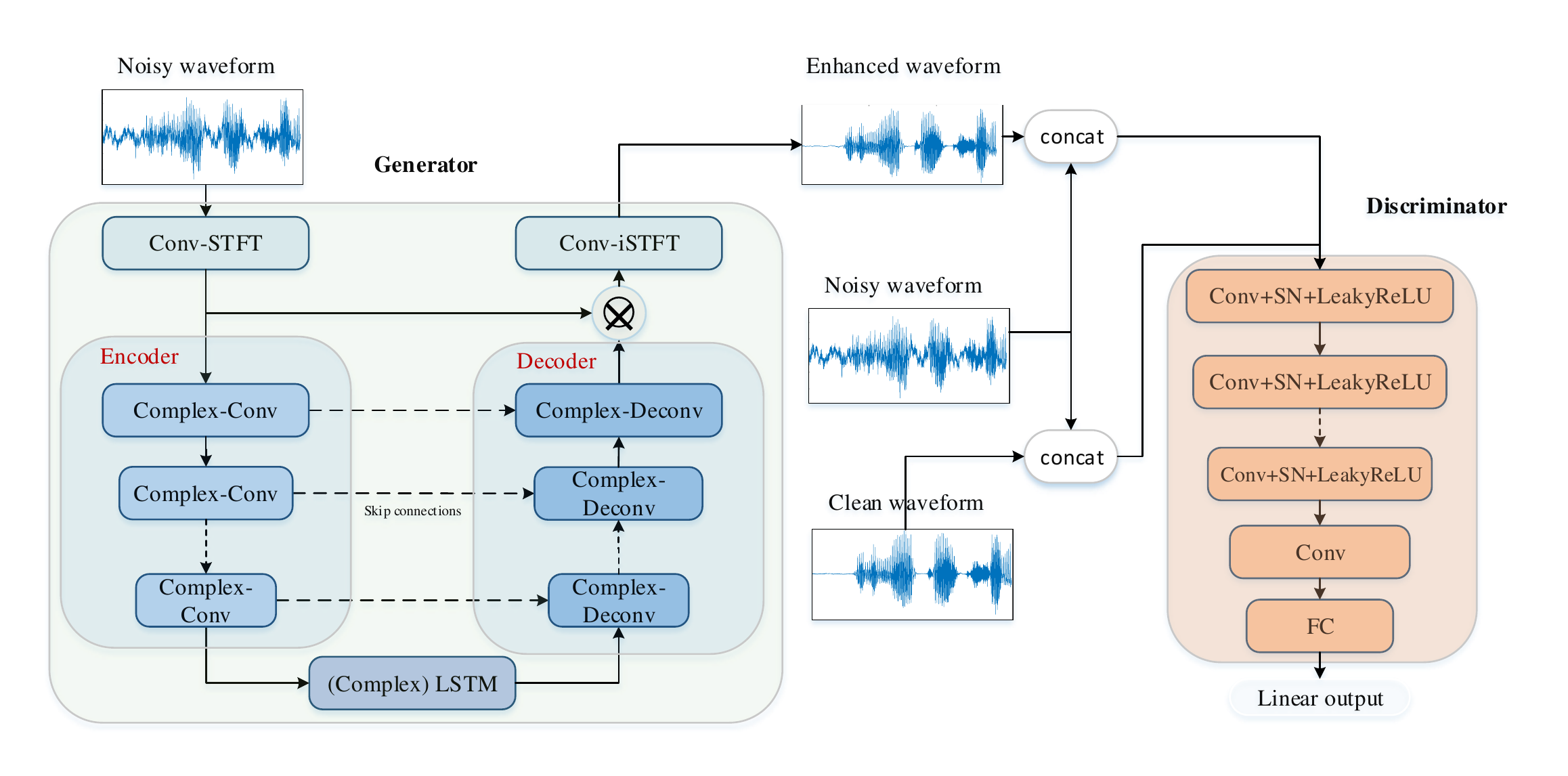}}
\caption{DCCRGAN structure.}
\label{fig2}
\end{figure}

The structure of the proposed DCCRGAN is shown in Fig. \ref{fig2}. In detail, G is an encoder-decoder network that regards the noisy waveform as inputs. After taking short time Fourier transform (STFT) by using a convolution layer, the high-level features are extracted by multiple complex convolution layers on encoder. Two LSTM layers between encoder and decoder are used to extract long-term information, which can be replaced by complex LSTM layers or complex BiLSTM layers. Decoder is a reversed process of the encoder. The skip connections are used to connect each encoding layer to the homologous decoding layer. Features dimension is recovered to the same size as the conv-STFT layer’s outputs after the last complex decoder layer. Decoder estimates the mask and applies to the inputs of encoder,   $\otimes$ represents element-wise multiplication. The estimated waveform is reconstructed through iSTFT operation by using a convolution layer.

\begin{figure}[t]
\centerline{\includegraphics[width=6cm]{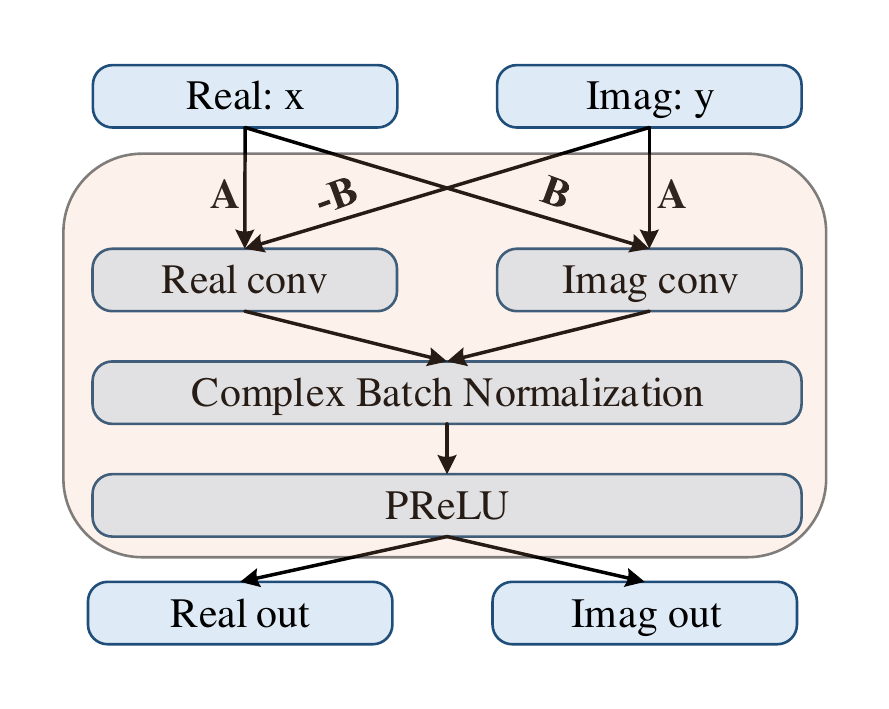}}
\caption{Complex module.}
\label{fig3}
\end{figure}

As shown in Fig. \ref{fig3}, each complex module is composed of a complex convolution layer, a complex batch normalization layer and a PReLU layer in encoder or decoder, except the last decoder layer. The complex convolution simulates the correlation between real and imaginary parts of the signal by complex multiplication. The complex convolution filter \textbf{W} and the complex input matrix  \textbf{H} is defined as $\textbf{W}=\textbf{A}+j\textbf{B}$ and $\textbf{H}=\textbf{X}+j\textbf{Y}$ respectively. The complex output features \textbf{V} of a complex convolution layer:
\begin{equation}
\begin{split}
\mathbf{V}=\mathbf{X}\ast\mathbf{W}=\left(\mathbf{A}\cdot\mathbf{X}-\mathbf{B}\cdot\mathbf{Y}\right)+j\left(\mathbf{B}\cdot\mathbf{X}+\mathbf{A}\cdot\mathbf{Y}\right),
\end{split}
\end{equation}
where \emph{j} represents imaginary part.

Similar to complex convolution, the complex LSTM can be defined as:
\begin{equation}
\begin{aligned}
\mathbf{V}_{rr}&=\text{LSTM}_r\left(\mathbf{X}\right),\mathbf{V}_{ir}=\text{LSTM}_i\left(\mathbf{X}\right)
\\
\mathbf{V}_{ri}&=\text{LSTM}_r\left(\mathbf{Y}\right),\mathbf{V}_{ii}=\text{LSTM}_i\left(\mathbf{Y}\right)
\\
\mathbf{V}&=\left(\mathbf{V}_{rr}-\mathbf{V}_{ii}\right)+j\left(\mathbf{V}_{ri}-\mathbf{V}_{ir}\right),
\end{aligned}
\end{equation}
where subscripts \emph{r} and \emph{i} represent the real and the imaginary parts of complex-value, respectively. \textbf{V} represents the outputs of complex LSTM layers.

The D network uses the analogous structure as SERGAN, but we apply spectral normalization (SN) in every layer of D, because the SN gets a better performance in controlling weight ranges of D than the gradient penalty according to \cite{9054256}. 

It’s difficult to train G by using original loss function directly. Previous researches \cite{DBLP:conf/interspeech/PascualBS17,8683799} showed that using L1 loss as an additional component is beneficial to the loss of G. L1 loss minimizes the distance between estimated waveform and clean waveform and is considered be more effective than L2 loss according to \cite{8462614}. Hence, the G loss is modified as:
\begin{equation}
\begin{split}
\mathcal{L}_G=-\mathbb{E}_{\left(\mathbf{x},\mathbf{y}\right)\sim \left(\mathcal{X},\mathcal{Y}\right)}\left[\log\left(\sigma\left(\text{D}\left(\text{G}\left(\mathbf{x}\right),\mathbf{x}\right)-\text{D}\left(\mathbf{y},\mathbf{x}\right)\right)\right)\right]\\
+\lambda_{L1}\lVert\text{G}\left(\mathbf{x}\right)-\mathbf{y}\rVert_1,
\end{split}
\end{equation}
where $\lambda_{L1}$ is a hyper parameter to control L1 component.

\section{EXPERIMENTS}

\subsection{Dataset}

We use the dataset released by Valentini et al. \cite{DBLP:conf/ssw/Valentini-Botinhao16} for fair comparison. The dataset is generated by simulation from two open data sources: Voice Bank corpus \cite{6709856} as speech data and DEMAND \cite{Joachim2013The} as environmental noise data. Voice Bank corpus is divided into a training set composed of 28 speakers and a test set composed of 2 speakers. To make the noisy training set, 40 kinds of noisy conditions are considered: 10 types of noise (2 artificial and 8 from the DEMAND database) and 4 kinds of signal-to-noise ratio (SNR) of 0, 5, 10, 15 dB, 11572 utterances in total. The test set including 5 types of noise conditions which are different from training noise conditions and another 4 SNR levels of 2.5, 7.5, 12.5 and 17.5 dB, 824 utterances in total. All utterances were down sample from 48 kHz to 16 kHz, and are sliced by a sliding window of length 16000 with 50\% overlap. We reconstruct the enhanced waveform by adding the estimated waveform with the same overlap and dividing the overlapped part by 2.

\subsection{Model setup}

Complex ratio mask (CRM) \cite{7364200} is estimated at decoder during the training process. Supposed that \textbf{X} and \textbf{Y} represent the complex-value STFT spectrum of noisy waveform and clean waveform, respectively, CRM can be defined as: $\text{CRM}=\left(\mathbf{X}_r\cdot\mathbf{Y}_r+\mathbf{X}_i\cdot\mathbf{Y}_i\right)~/~\left(\mathbf{X}_r^{2}+\mathbf{X}_i^{2}\right)+j\left(\mathbf{X}_r\cdot\mathbf{Y}_i-\mathbf{X}_i\cdot\mathbf{Y}_r\right)~/~\left(\mathbf{X}_r^{2}+\mathbf{X}_i^{2}\right)$, where subscript \emph{r} and \emph{i} denote the real and imaginary parts of the complex spectrum, respectively. CRM can also perform a rotation on a polar coordinates and correct phase errors \cite{DBLP:conf/iclr/ChoiKHKHL19}, the representation $\hat{\mathbf{M}}=\hat{\mathbf{M}}_r+j\hat{\mathbf{M}}_i$ is expressed in polar coordinates as: $\hat{\mathbf{M}}_{mag}=\sqrt{\hat{\mathbf{M}}_r^{2}+\hat{\mathbf{M}}_i^{2}}$, $\hat{\mathbf{M}}_{phase}=\arctan2\left(\hat{\mathbf{M}}_r,\hat{\mathbf{M}}_i\right)$.

Estimated complex spectrogram $\hat{\mathbf{X}}$ can be calculated as:
\begin{equation}
\begin{aligned}
\hat{\mathbf{X}}_{CRM}&=\left(\mathbf{X}_r\cdot\hat{\mathbf{M}}_{r}-\mathbf{X}_i\cdot\hat{\mathbf{M}}_{i}\right)+j\left(\mathbf{X}_r\cdot\hat{\mathbf{M}}_{i}+\mathbf{X}_i\cdot\hat{\mathbf{M}}_{r}\right)
\\
\hat{\mathbf{X}}_{polar}&=\mathbf{X}_{mag}\cdot\hat{\mathbf{M}}_{mag}\cdot\exp\left(\mathbf{X}_{phase}\cdot\hat{\mathbf{M}}_{phase}\right).
\end{aligned}
\end{equation}

Besides, we can estimate the mask of the real and imaginary parts of $\hat{\mathbf{X}}$, respectively. Therefore $\hat{\mathbf{X}}$ can be defined as:
\begin{equation}
\begin{aligned}
\hat{\mathbf{X}}_{real}=\mathbf{X}_r\cdot\hat{\mathbf{M}}_{r}+j\left(\mathbf{X}_i\cdot\hat{\mathbf{M}}_{i}\right).
\end{aligned}
\end{equation}

Estimated waveform is reconstructed through iSTFT operation by using the conv-iSTFT layer. We used DCCRGAN-C, DCCRGAN-E and DCCRGAN-R represent the masking method of $\hat{\mathbf{X}}_{CRM}$, $\hat{\mathbf{X}}_{polar}$ and $\hat{\mathbf{X}}_{real}$, respectively.

We set the window length and hop size to 25 ms and 6.25 ms, respectively, and a 512 FFT length in the conv-STFT layer and the conv-iSTFT layer in G network. There are 6 complex convolution layers in encoder and the channel number is \{16, 32, 64, 128, 256, 256\}. The stride and the kernel size are set to (2, 1) and (5, 2) in each layer, respectively. The real LSTM layers consist of 2 layers with 256 units meanwhile the complex LSTM and the complex BiLSTM are 2 layers with 128 units for the real and imaginary parts, respectively. The channel number of decoder follows the reverse setting of the encoder. In the last layer of decoder, none complex batch normalization and activation function are used during training DCCRGAN-C and DCCRGAN-R, while a $\tanh$ non-linear function is used to limit the mask magnitude for training DCCRGAN-E.

The D network used in DCCRGAN is similar to SERGAN, which has two input channels, one for clean waveform \textbf{y} or estimated waveform G(\textbf{x}), and one for noisy waveform \textbf{x}. The structure is made up of 11 one-dimensional convolution layers and the channel number is \{16, 32, 32, 64, 64, 128, 128, 256, 256, 512, 1024\} with filter-length 31 and stride 2 in each layer. Spectral normalization and LeakyReLU with $\alpha=0.3$ are applied after every one-dimensional layer. There is an additional one-dimensional convolution layer with filter-length 1 and stride 1 after the last activation layer, and the output is fed into a fully connection layer with linear activation to perform a binary classification.

We trained all the models using Adam optimizer \cite{DBLP:journals/corr/KingmaB14} for 100 epochs and a batch size of 64, the learning rate is set to 0.001 and it will decay 0.5 when the loss goes up. The hyper parameter $\lambda_{L1}$ is set to 100 according to \cite{DBLP:conf/interspeech/PascualBS17}.

\subsection{Experimental result}

Several objective metrics are adopted to evaluate the speech enhancement performance: PESQ \cite{941023}, CSIG \cite{4389058}, CBAK \cite{4389058}, COVL \cite{4389058} and STOI \cite{DBLP:conf/icassp/TaalHHJ10}. Higher scores indicate a better performance for all metrics.

\begin{table}[t]
\caption{Comparisons of DCCRGAN with different mask and different recurrent layers by using relativistic loss.}
\begin{center}
\begin{tabular}{c|ccccc}
\hline
Setting & PESQ & CSIG & CBAK & COVL & STOI\\
\hline
\multicolumn{6}{c}{DCCRGAN with LSTM layers}\\
\hline
DCCRGAN-R&2.74	 &3.89&3.4&3.28&0.945\\
DCCRGAN-E&2.72&3.94&3.42&3.29&0.946\\
DCCRGAN-C&2.79&3.97&3.46&3.36&0.948\\
\hline
\multicolumn{6}{c}{DCCRGAN with complex LSTM layers}\\
\hline
DCCRGAN-R&2.78	 &3.89&3.44&3.32&0.946\\
DCCRGAN-E&2.76&3.83&3.42&3.26&0.945\\
DCCRGAN-C&2.77&\textbf{3.99}&3.41&\textbf{3.37}&0.945\\
\hline
\multicolumn{6}{c}{DCCRGAN with complex BiLSTM layers}\\
\hline
DCCRGAN-R&2.79	 &3.91&3.42&3.32&0.946\\
DCCRGAN-E&2.76&3.85&3.43&3.25&0.947\\
DCCRGAN-C&\textbf{2.83}&3.94&\textbf{3.50}&3.35&\textbf{0.949}\\
\hline
\end{tabular}
\label{tab1}
\end{center}
\end{table}

Table \ref{tab1} shows the speech enhancement performance of different mask methods with different LSTM layers. DCCRGAN-R and DCCRGAN-E achieve similar performance in different LSTM layers. DCCRGAN-C achieves higher performance in all metrics except using complex LSTM layers. We comprehensively analyze the influence of different recurrent layers on generator network. Although some metrics result of complex LSTM layers are lower than real LSTM layers (e.g. PESQ and STOI in DCCRGAN-C), the complex LSTM layers and complex BiLSTM layers tend to reach a better performance. All scores of complex recurrent modules are higher than that of real LSTM layers, which indicates that the speech intelligibility and quality can be significantly improved by using complex recurrent modules.

Table \ref{tab2} compares the performance of proposed models using complex BiLSTM layers and SOTA GAN-based SE systems. SERGAN \cite{8683799}, CRGAN \cite{DBLP:conf/interspeech/ZhangDSWSZSL20} and CP-GAN \cite{9054060} adopted U-Net, convolution recurrent network (CRN) and densely-connected in G, respectively. We can see that relativistic average loss performs better than relativistic loss in the same G structure, CP-GAN modifies the network via densely-connected feature pyramid generator, and achieves a remarkable improvement of the metric score except PESQ value. The proposed DCCRGAN achieves better performance compared to other G structures under the same objective function, especially a dramatic increase in CBAK, COVL and STOI. The results illustrate that the phase-aware complex convolution operation considers the correlation between real and imaginary parts of the signal, and can achieve a better performance than other structures. Furthermore, we compare the GAN-based model using relativistic loss with other GAN-based systems with different objective function. All the GAN-based systems improve speech quality over noisy waveform. The GAN-based systems using relativistic loss and relativistic average loss achieve better noise reduction performance than other GAN-based systems. The proposed DCCRGAN outperforms other SOTA GAN-based speech enhancement system for all metrics.

\begin{table}[t]
\caption{Comparison of proposed DCCRGAN and SOTA GAN-based systems. R(a)-SGAN and Metric-GAN refers the result adopted from \cite{DBLP:conf/interspeech/ZhangDSWSZSL20}. CP-GAN is trained by using relativistic loss function. The prefix R and Ra of the DCCRGAN strand for relativistic loss and relativistic average loss, respectively.}
\begin{center}
\begin{tabular}{c|ccccc}
\hline
Setting & PESQ & CSIG & CBAK & COVL & STOI\\
\hline
Noisy&1.97&3.35&2.44&2.63&0.921\\
SEGAN \cite{DBLP:conf/interspeech/PascualBS17}&2.16&3.48	&2.94&2.80&0.93\\
MMSE-GAN \cite{8462068}&2.53&3.80&3.12&3.15&0.93\\
Metric-GAN \footnotemark[1]\cite{DBLP:conf/icml/FuLTL19}&2.49&3.81&3.05&3.13&0.925\\
\hline
\multicolumn{6}{c}{GAN-based model with relativistic loss}\\
\hline
R-SGAN \cite{8683799}&2.51&3.82&3.16&3.15&0.937\\
Ra-SGAN \cite{8683799}&2.57&3.83&3.28&3.20&0.937\\
R-CRGAN \cite{DBLP:conf/interspeech/ZhangDSWSZSL20}&2.72&3.67&3.09&3.17&0.932\\
Ra-CRGAN \cite{DBLP:conf/interspeech/ZhangDSWSZSL20}&2.81&3.72&3.16&3.25&0.936\\
CP-GAN \cite{9054060}&2.64&3.93&3.29&3.28&0.94\\
\hline
\multicolumn{6}{c}{DCCRGAN with complex BiLSTM layers}\\
\hline
R-DCCRGAN-R&2.79&3.91&3.42&3.32&0.946\\
R-DCCRGAN-E&2.76&3.85&3.43&3.25&0.947\\
R-DCCRGAN-C&\textbf{2.83}&3.94&\textbf{3.50}&3.35&\textbf{0.949}\\
Ra-DCCRGAN-R&\textbf{2.83}&3.99&\textbf{3.50}&3.38&0.948\\
Ra-DCCRGAN-E&2.77&3.86&3.46&3.31&\textbf{0.949}\\
Ra-DCCRGAN-C&2.82&\textbf{4.01}&3.48&\textbf{3.40}&\textbf{0.949}\\
\hline
\end{tabular}
\label{tab2}
\end{center}
\end{table}
\footnotetext[1]{The score refered from [21] is different from the ORIGINAL score in [38], due to the different training epochs between [21] and [38].}

\section{conclusions}

In this paper, we proposed a GAN-based SE system which uses deep complex convolution recurrent structure in generator network. We trained DCCRGAN in an end-to-end way to prevent phase mismatch problem occurring in the T-F domain. A better performance can be obtained by using complex BiLSTM layers under the proposed DCCRGAN framework. Further results demonstrate that the complex-value multiplication operation can achieve a better performance than other generator network with the same objective function, in addition to improving the speech enhancement performance when comparing with SOTA GAN-based SE systems.

\bibliographystyle{IEEEtran}
\bibliography{ref}
\end{document}